\documentclass[conference, a4paper]{IEEEtran}

\usepackage{algorithm,algorithmic}
\usepackage{amsmath,amssymb}
\usepackage{array}
\usepackage{balance}
\usepackage{booktabs}
\usepackage{cite}
\usepackage{color}
\usepackage{epsfig}
\usepackage{easyReview}
\usepackage{fancyhdr}
\usepackage{float}
\usepackage[nomain,acronym,automake,toc,nopostdot]{glossaries}
\usepackage{graphicx}
\usepackage[hidelinks]{hyperref}
\usepackage{mathrsfs}
\usepackage{mathtools}
\usepackage{multirow}
\usepackage{psfrag}
\usepackage{stfloats}
\usepackage{subfigure}
\usepackage[left=1.44cm,right=1.44cm,top=2.88cm]{geometry}

\makeglossaries
\newacronym{1d}{1D}{one-dimensional}
\newacronym{2d}{2D}{two-dimensional }
\newacronym{3d}{3D}{three-dimensional}
\newacronym{3gpp}{3GPP}{3rd Generation Partnership Project}
\newacronym{5g}{5G}{fifth-generation}
\newacronym{6g}{6G}{sixth-generation}

\newacronym{ap}{AP}{access point}
\newacronym{ai}{AI}{artificial intelligence}
\newacronym{amp}{AMP}{approximate message passing}
\newacronym{ask}{ASK}{amplitude shift keying}
\newacronym{awgn}{AWGN}{additive white Gaussian noise}

\newacronym{ber}{BER}{bit error rate}
\newacronym{bfgs}{BFGS}{Broyden–Fletcher–Goldfarb–Shanno}
\newacronym{bler}{BLER}{bit error rate}
\newacronym{bp}{BP}{belief propagation}
\newacronym{bpsk}{BPSK}{binary phase shift keying}
\newacronym{bs}{BS}{base station}

\newacronym{cbsm}{CBSM}{correlation-based stochastic model}
\newacronym{csirs}{CSI-RS}{channel state information reference signal}
\newacronym{cdf}{CDF}{cumulative distribution function}
\newacronym{cdma}{CDMA}{code division multiple access}
\newacronym{cfr}{CFR}{channel frequency response}
\newacronym{cg}{CG}{conjugate gradient descent}
\newacronym{clt}{CLT}{central limit theorem}
\newacronym{cots}{COTS}{commercial off-the-shelf}
\newacronym{csi}{CSI}{channel state information }
\newacronym{cvp}{CVP}{closest vector problem}

\newacronym{dof}{DoF}{degrees of freedom}

\newacronym{edw}{EDW}{exponentially decaying window}
\newacronym{elaa}{ELAA}{extremely large aperture array}
\newacronym{embb}{eMBB}{enhanced mobile broadband}
\newacronym{emtc}{eMTC}{enhanced machine type communication}
\newacronym{etsi}{ETSI}{European Telecommunications Standards Institute}

\newacronym{ff}{FF}{far-field}
\newacronym{fcsd}{FCSD}{fixed-complexity sphere decoder}
\newacronym{fec}{FEC}{forward error correction}
\newacronym{fspl}{FSPL}{free space path loss}

\newacronym{gbsm}{GBSM}{geometry-based stochastic model}
\newacronym{gd}{GD}{gradient descent}
\newacronym{gmsk}{GMSK}{Gaussian minimum shift keying}
\newacronym{gs}{GS}{Gauss-Seidel}
\newacronym{gsm}{GSM}{global system for mobile communication}

\newacronym{iid}{i.i.d.}{independently and identical distributed}
\newacronym{ils}{ILS}{integer least-squares}
\newacronym{ind}{i.n.d.}{independently and non-identical distributed}
\newacronym{imt}{IMT}{International Mobile Telecommunications}
\newacronym{isac}{ISAC}{integrated sensing and communication}
\newacronym{isi}{ISI}{intersymbol interference}
\newacronym{itur}{ITU-R}{International Telecommunication Union Radiocommunication Sector}
\newacronym{iui}{IUI}{inter-user interference}

\newacronym{ji}{JI}{Jacobi iteration}
\newacronym{jsac}{JSAC}{joint sensing and communication}

\newacronym{las}{LAS}{likelihood ascent search}
\newacronym{lbfgs}{LBFGS}{limited-memory Broyden–Fletcher–Goldfarb–Shanno}
\newacronym{lll}{LLL}{Lenstra-Lenstra-Lov\'{a}sz}
\newacronym{llr}{LLR}{log-likelihood ratio}
\newacronym{los}{LoS}{line of sight}
\newacronym{lr}{LR}{lattice reduction}
\newacronym{lmmse}{LMMSE}{minimum mean square error}
\newacronym{lsd}{LSD}{list sphere decoder}
\newacronym{lte}{LTE}{long-term evolution}

\newacronym{map}{MAP}{maximum a posteriori}
\newacronym{mf}{MF}{matched filter}
\newacronym{mimo}{MIMO}{multiple-input multiple-output}
\newacronym{mld}{MLD}{maximum likelihood detection}
\newacronym{mlsd}{MLSD}{maximum likelihood sequence detection}
\newacronym{mmimo}{mMIMO}{massive multiple-input multiple-output}
\newacronym{mrc}{MRC}{maximum ratio combining}
\newacronym{mse}{MSE}{mean square error}
\newacronym{ms}{MS}{matrix-splitting}
\newacronym{mt}{MT}{mobile terminal}

\newacronym{nf}{NF}{near-field}
\newacronym{nlos}{NLoS}{non-LoS}
\newacronym{nic}{NIC}{network interface card}

\newacronym{od}{OD}{orthogonality defect}
\newacronym{ofdm}{OFDM}{orthogonal frequency division multiplexing}
\newacronym{ofdma}{OFDMA}{orthogonal frequency division multiple access}

\newacronym{pdf}{PDF}{probability distribution function}
\newacronym{pda}{PDA}{probabilistic data association}
\newacronym{pep}{PEP}{pairwise error probability}
\newacronym{pl}{PL}{path-loss}
\newacronym{pmf}{PMF}{probability mass function}
\newacronym{pwm}{PWM}{plane-wave model}

\newacronym{qam}{QAM}{quadrature amplitude modulation}
\newacronym{qpsk}{QPSK}{quadrature phase shift keying}
\newacronym{qn}{QN}{quasi-Newton}

\newacronym{rts}{RTS}{reactive tabu search}
\newacronym{ri}{RI}{Richardson iteration}
\newacronym{ris}{RIS}{reconfigurable intelligent surface}
\newacronym{rss}{RSS}{received signal strength}
\newacronym{rssi}{RSSI}{received signal strength indicator}
\newacronym{rx}{Rx}{receiver}
\newacronym{rzf}{RZF}{regularized-ZF}

\newacronym{sa}{SA}{Seysen's algorithm}
\newacronym{sd}{SD}{steepest descent}
\newacronym{sdr}{SDR}{semidefinite relaxation}
\newacronym{sdr2}{SDR}{software-defined radio}
\newacronym{ser}{SER}{symbol error rate}
\newacronym{sf}{SF}{shadow fading}
\newacronym{sic}{SIC}{succesive interference cancellation}
\newacronym{sinr}{SINR}{signal-to-interference-plus-noise ratio}
\newacronym{siso}{SISO}{single input single output}
\newacronym{snr}{SNR}{signal-to-noise ratio}
\newacronym{sns}{SNS}{spatial non-stationarity}
\newacronym{sota}{SoTA}{state-of-the-art}
\newacronym{ssor}{SSOR}{symmetric successive over-relaxation}
\newacronym{svd}{SVD}{singular value decomposition }
\newacronym{swm}{SWM}{spherical-wave model}

\newacronym{tr}{TR}{Technical Report}
\newacronym{ts}{TS}{tabu search}
\newacronym{tx}{Tx}{transmitter}

\newacronym{uca}{UCA}{uniform cylindrical array}
\newacronym{ue}{UE}{user equipment}
\newacronym{ula}{ULA}{uniform linear array}
\newacronym{ura}{URA}{uniform rectangular array}
\newacronym{uma}{UMa}{urban macro}
\newacronym{umi}{UMi}{urban micro}
\newacronym{upa}{UPA}{uniform planar array}
\newacronym{usrp}{USRP}{universal software radio peripheral}

\newacronym{vblast}{V-BLAST}{vertical Bell Labs layered space-time}

\newacronym{wifi}{Wi-Fi}{wireless fidelity}

\newacronym{xlmimo}{XL-MIMO}{extra-large multiple-input multiple-output}
\newacronym{xr}{XR}{extended reality}

\newacronym{zf}{ZF}{zero-forcing}

\definecolor{sblue}{RGB}{0,51,120}
\definecolor{sred}{RGB}{200,51,130}

\newcommand{\figref}[1]{Fig. \ref{#1}}

\newcommand{\secref}[1]{Section \ref{#1}}
\renewcommand{\eqref}[1]{(\ref{#1})}

\ifCLASSINFOpdf
\else
\fi

\begin{document}
\title{Wi-Fi Beyond Communications: \\ Experimental Evaluation of Respiration Monitoring and Motion Detection Using COTS Devices}
\author{Jiuyu Liu, Yi Ma, and Rahim Tafazolli\\
{\small 5GIC and 6GIC, Institute for Communication Systems, University of Surrey, Guildford, UK, GU2 7XH}\\
{\small Emails: (jiuyu.liu, y.ma, r.tafazolli)@surrey.ac.uk}}
\markboth{}%
{}

\maketitle

\begin{abstract}
Wi-Fi sensing has become an attractive option for non-invasive monitoring of human activities and vital signs.
This paper explores the feasibility of using state-of-the-art commercial off-the-shelf (COTS) devices for Wi-Fi sensing applications, particularly respiration monitoring and motion detection.
We utilize the Intel AX210 network interface card (NIC) to transmit Wi-Fi signals in both 2.4 GHz and 6 GHz frequency bands.
Our experiments rely on channel frequency response (CFR) and received signal strength indicator (RSSI) data, which are processed using a moving average algorithm to extract human behavior patterns.
The experimental results demonstrate the effectiveness of our approach in capturing and representing human respiration and motion patterns.
Furthermore, we compare the performance of Wi-Fi sensing across different frequency bands, highlighting the advantages of using higher frequencies for improved sensitivity and clarity.
Our findings showcase the practicality of using COTS devices for Wi-Fi sensing and lay the groundwork for the development of non-invasive, contactless sensing systems.
These systems have potential applications in various fields, including healthcare, smart homes, and Metaverse.
\end{abstract}
\glsresetall
\glsresetall

\section{Introduction} \label{sec01}
The Metaverse, a concept that envisions the seamless integration of the physical and virtual worlds, is expected to become a reality with the advancement of future wireless network systems that provide both high data rates and sensing capabilities \cite{Zhou2023, Liu2024, Qiao2024}.
\Gls{wifi} has become ubiquitous, enabling high-speed internet access in various settings, and has emerged as a promising technology for enabling the Metaverse vision.
This technique capitalizes on the measurable changes in the wireless channel (e.g., \gls{csi}, \gls{cfr}, and \gls{rssi}) caused by movements and physiological activities \cite{Ma2021, Li2023, Gao2023}.
By leveraging these changes, \gls{wifi} sensing allows for continuous monitoring without the need for direct contact or wearable devices \cite{Dai2023}.
To facilitate the widespread adoption and implementation of \gls{csi}-based sensing in the Metaverse, the development and deployment of cost-effective hardware solutions that support the required functionalities are essential.

\Gls{sdr2} devices, such as \gls{usrp}, are capable of transmitting \gls{wifi} pilot signals and extracting high-resolution channel data \cite{Jiang2022}. 
Although this data can be utilized to achieve high sensing performance, the cost of these devices is prohibitive for practical applications \cite{Liu2016}.
Therefore, this paper focuses on \gls{cots} devices for practical \gls{wifi} sensing applications.
Recent studies have demonstrated the ability of \gls{cots} \gls{wifi} devices to monitor human respiration \cite{Gu2018, Lai2022a, Zhang2023b}. 
These works employ the Intel 5300 \gls{nic} with specific software to measure the \gls{cfr} and capture periodic respiratory behavior.
Real-time signal processing is possible, but with a delay of several respiration cycles due to the need to wait for the periodicity to occur.
However, compared to the \gls{rssi} data, which is available to all the \gls{wifi} devices, \gls{cfr} data is not accessible for most devices.
Furthermore, experiments comparing the accuracy between $2.4$ and $5$ $\mathrm{GHz}$ \gls{wifi} signals have shown that higher frequency bands are expected to further enhance the sensing precision and resolution \cite{Guan2023}.

Current experimental works on \gls{wifi} sensing primarily focus on measuring cyclically stationary phenomena. 
However, the vast majority of human activities, such as people moving from one location to another, are not cyclical \cite{Zhang2024}.
Moreover, it is important to note that even respiration does not always exhibit cyclically stationary behavior.
In reality, respiration patterns can be highly irregular, with people holding their breath voluntarily or experiencing involuntary pauses in breathing, such as during sleep apnea, which can be a serious health concern \cite{Korkalainen2021}.
These situations need to be detected promptly, as even a ten-second delay may be too late.
Furthermore, the experiments mentioned above are conducted using Wi-Fi 4 or 5 standards \cite{Guan2023}. 
The latest \gls{wifi} \gls{nic} is based on the IEEE 802.11be (i.e., Wi-Fi 7) standard, which includes the $6$ $\mathrm{GHz}$ frequency band. However, to the best of our knowledge, there are no published CSI-based sensing experiments using \gls{wifi} 7 signals. 
All the concerns above motivate this paper.

In this paper, we employ the Intel AX210 \gls{nic}, one of the most advanced \gls{cots} \gls{wifi} devices.
This device supports $2.4$, $5$, and $6$ $\mathrm{GHz}$ frequency bands and is capable of transmitting \gls{wifi} 7 signals \cite{Jiang2022}.
We conduct two types of experiments to capture human behavior: respiration monitoring and motion detection.
In these experiments, we utilize both \gls{cfr} and \gls{rssi}.
While \gls{cfr} is specific to certain devices, \gls{rssi} is accessible to all \gls{wifi} devices.
Unlike previous works, our participant does not breathe or move at a fixed frequency, which poses a challenge in capturing these behaviors. 
To address this issue, we propose the use of a moving average algorithm, which is capable of capturing respiration and movement with only millisecond-level delay.
Our experimental results demonstrate that both \gls{cfr} and \gls{rssi} can effectively track respiration and motion behaviors. 
Furthermore, we observe that higher frequencies can provide more apparent fluctuations in the captured data.
These findings suggest that the use of higher frequency bands, such as those available in \gls{wifi} 7, has the potential to improve the performance of \gls{wifi} sensing systems.

\section{System Description and Problem Statement} \label{sec02}
In this section, we first introduce the mathematical definitions of \gls{cfr} and \gls{rssi}.
Following these definitions, we present the problem statement that motivates our research.

\subsection{System Description}
\subsubsection{CFR}
In \gls{wifi} systems, the transmitted symbols are \gls{ofdm} symbols.
Denote $H(f_{k}, t)$ as the \gls{cfr} of the $k^{th}$ sub-carrier at the center frequency $f_{k}$ and time slot $t$, it can be expressed as follows
\begin{equation}
	H(f_{k}, t) = \|H(f_{k}, t)\| e^{-j \angle H(f_{k}, t)},
\end{equation}
where $\|H(f_{k}, t)\|$ and $\angle H(f_{k}, t)$ represent the amplitude and phase of the $k^{th}$ sub-carrier at time slot $t$, respectively.
In this work, we focus on the amplitude of each sub-carrier. 
To facilitate our presentation, we define it as follows
\begin{equation} \label{eqn02}
	A_{k,t} \triangleq \|H(f_{k},t)\|.
\end{equation}
As discussed in \secref{sec01}, previous works have utilized \gls{wifi} signals to sense human activities with cyclic stationary behavior.
To sense such activities, they usually transform $A_{k,t}$ (e.g., using Fourier transform) into the frequency domain and apply a suitable frequency-domain filter to isolate the relevant frequency components.

\subsubsection{RSSI}
In wireless communications, \gls{rssi} is commonly used to measure the signal strength.
Denote $P_{t}$ to be the received power in $\mathrm{dBm}$ at the receiver at the $t^{th}$ time slot, RSSI is defined as follows
\begin{equation} \label{eqn03}
	I_{t} = \mathcal{Q} \left(10 \cdot \log_{10} \left( \frac{P_{t}}{P_{\textsc{ref}}} \right) \right),
\end{equation}
where $P_{\textsc{ref}}$ is the reference power level, typically set to $1$ $\mathrm{mW}$, which corresponds to $0$ $\mathrm{dBm}$.
$\mathcal{Q}(\cdot)$ is a quantization function that quantizes the resolution of RSSI to $1$ $\mathrm{dB}$.
It provides a measure of the signal strength at the receiver, with higher values indicating a stronger received signal.

Unlike \gls{cfr}, which is specific to certain devices, RSSI is accessible to all \gls{wifi} devices.
This makes RSSI a more universally available metric for \gls{wifi} sensing applications.
However, RSSI has a lower resolution compared to \gls{cfr}, as it is quantized to $1$ $\mathrm{dB}$ steps.

\subsection{Problem Statement} \label{sec02b}
Despite the progress made in utilizing \gls{cfr} for sensing cyclic stationary human activities, several challenges still need to be addressed:
\textit{1)} Most human activities do not exhibit cyclic stationary behavior, making it difficult to capture these activities using the frequency-domain filtering approach mentioned in the previous subsection.
\textit{2)} Previous works have primarily focused on \gls{wifi} 4/5 signals, and there is a lack of experimental studies using the latest \gls{wifi} 7 signals, which include the $6$ $\mathrm{GHz}$ frequency band.
\textit{3)} \Gls{cfr}-based sensing requires specific professional equipment and software, restricting its widespread adoption in practical applications.
\Gls{rssi} is accessible to all \gls{wifi} devices but its potential for sensing non-cyclic human activities and performance compared to \gls{cfr} across different frequency bands remains to be investigated.
All these challenges motivate the following of this paper.

\section{Hardware Specification, Experimental Design and Signal Processing Algorithm} \label{sec03}
This section provides a detailed specification of the hardware used in our experiments, presents an overview of the experimental design and procedures, and introduces the moving average algorithm used to process the measured data.

\subsection{Specifications of Hardware}
\begin{table}[]
	\caption{Hardware Details and Operating Frequency}
	\label{tab01}
	\centering
	\renewcommand{\arraystretch}{1.3}
	\resizebox{0.35\textwidth}{!}{
	\begin{tabular}{|cc|}
		\hline
		\multicolumn{2}{|c|}{Experiment Setup}                                                                              \\ \hline
		\multicolumn{1}{|c|}{Environment}                    & Indoor Office                                                \\ \hline
		\multicolumn{1}{|c|}{Type of \gls{wifi} NIC}              & Intel AX210                                                  \\ \hline
		\multicolumn{1}{|c|}{Number of Tx antennas} & $1$                                                          \\ \hline
		\multicolumn{1}{|c|}{Number of Rx antennas}    & $4$                                                          \\ \hline
		\multicolumn{1}{|c|}{Carrier frequency}              & $2.4$ $\mathrm{GHz}$, $6$ $\mathrm{GHz}$ \\ \hline
		\multicolumn{1}{|c|}{Bandwidth}                      & $20$ $\mathrm{MHz}$                                          \\ \hline
		\multicolumn{1}{|c|}{Number of sub-carriers}         & $245$                                                        \\ \hline
	\end{tabular}
	}
	\vspace{-1em}
\end{table}

In our experiments, we utilized the Intel AX210, one of the most advanced Wi-Fi devices available.
The AX210 \gls{nic} supports multiple carrier frequencies, including $2.4$ $\mathrm{GHz}$, $5$ $\mathrm{GHz}$, and $6$ $\mathrm{GHz}$, enabling a comprehensive analysis of signal behaviors across different bands.
Thanks to the PicoSenses software, AX210 can be used to transmit \gls{wifi} 7 signals using the IEEE 802.11be standard \cite{Jiang2022}.
For our \gls{tx}, we utilized a Dell G5 laptop equipped with one AX210 \gls{nic}. The laptop runs on Ubuntu 20.04 operating system and features an Intel Core i7 10750H $2.60$ $\mathrm{GHz}$ CPU and $16$ $\mathrm{GB}$ of memory.
On the \gls{rx} side, we used an assembled desktop with two AX210 \glspl{nic}.
The desktop is powered by an Intel i5 8400 CPU running at $4$ $\mathrm{GHz}$ and has $16$ $\mathrm{GB}$ of memory.
The desktop runs on Ubuntu 22.04 operating system.

To ensure a fair comparison across all frequency bands, our experiment utilizes a bandwidth of $20$ $\mathrm{MHz}$.
This choice was made because the $2.4$ $\mathrm{GHz}$ band is legally restricted to a maximum bandwidth of $20$ $\mathrm{MHz}$. 
At this bandwidth, the system processes a total of $245$ sub-carriers.
Table \ref{tab01} presents a comprehensive overview of the key hardware components and operating parameters used in our experiments, providing a clear understanding of the experimental setup.

\subsection{Experimental Design}
This work focuses on two distinct experiments: respiration monitoring and motion detection.
Each experiment requires a specific setup with different transceiver locations.
In the following two subsections, we provide a detailed description of each experiment, including the precise placement of the transceivers and the instructions given to the participants.

\subsubsection{Respiratory Monitoring}
\begin{figure}[t]
	\centering
	\subfigure[\label{figLay1} Experimental Layout]{
		\begin{minipage}[t]{0.48\textwidth}	
			\centering
			\includegraphics[width=6.6cm]{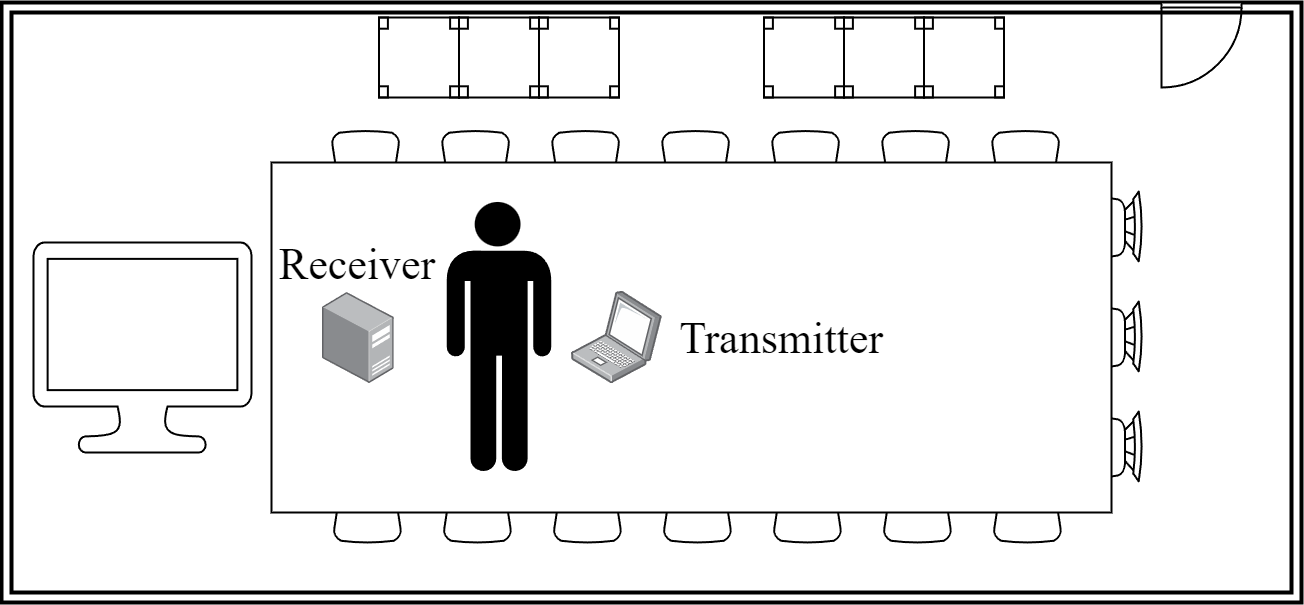}
	\end{minipage}}
	\subfigure[\label{figEnv1} Experimental Environment]{
		\begin{minipage}[t]{0.48\textwidth}
			\centering
			\includegraphics[width=6.6cm]{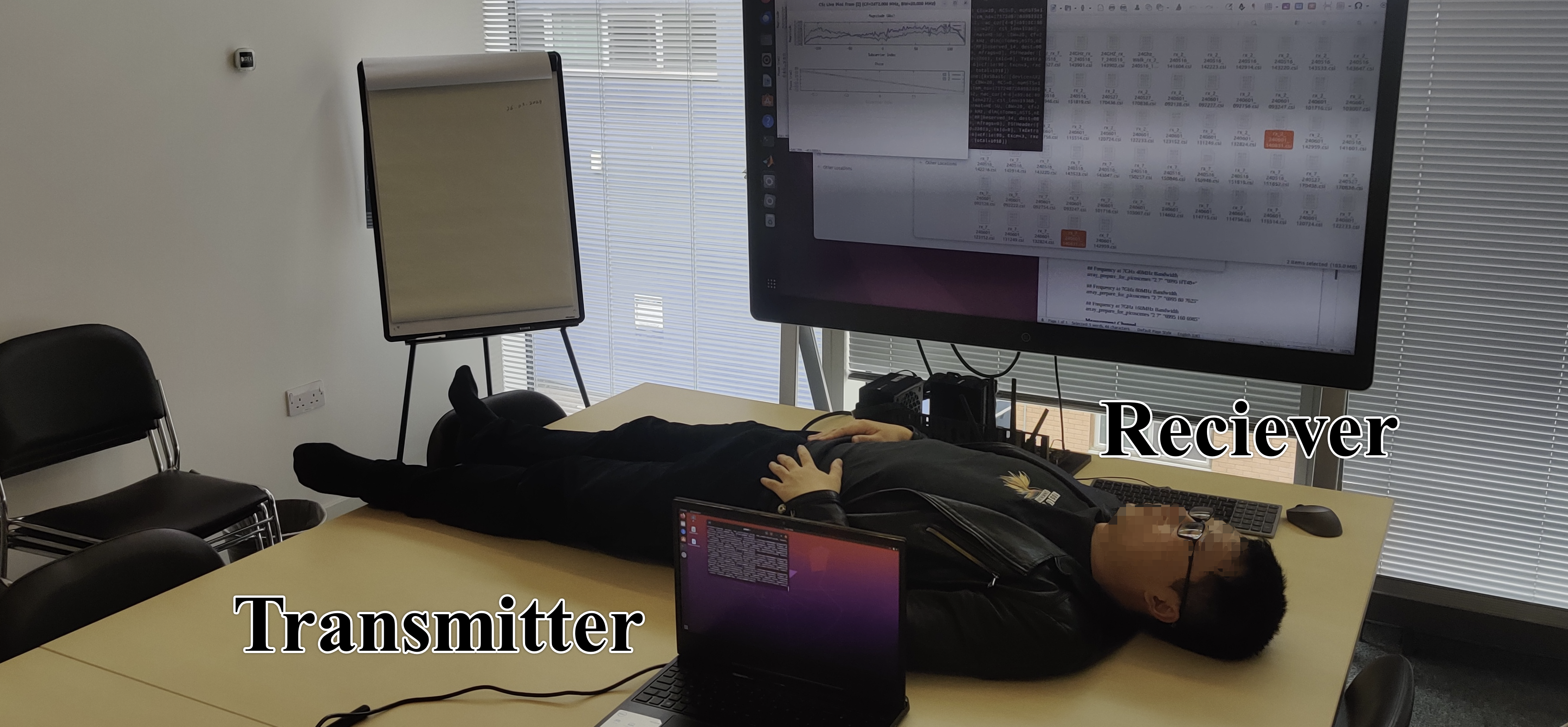}  
	\end{minipage}}
	\caption{\label{figExp1} The experimental layout and environment of respiration monitoring.}
	\vspace{-1em}
\end{figure}
In the respiration monitoring experiment, our primary objective is to capture the breathing patterns of a stationary participant.
\figref{figLay1} illustrates the layout of the experiment, where the participant lies down between the \gls{tx} and \gls{rx}, which are separated by a distance of approximately $2$ meters.
The transmitter sends Wi-Fi signals across the room, while the receiver captures the reflected signals.
As the participant breathes, their chest rises and falls, creating subtle changes in the signal propagation path between the \gls{tx} and \gls{rx}.
\figref{figEnv1} shows the actual experimental environment, where a person lies down on their back on a mat or a low platform.
The power of \gls{los} links is significantly greater than that of \gls{nlos} links \cite{Liu2021, Liu2023c}.
This position allows for clear monitoring of the person's chest movements during respiration.

The experimental setup is designed to capture the variations in the wireless signal caused by the participant's chest movements during inhalation and exhalation.
As the person inhales and exhales, their chest rises and falls, creating slight variations in the reflected Wi-Fi signals.
These variations can be detected and processed by the receiver to extract information about the person's respiratory rate and patterns.
Moreover, different from previous works, our objective is not only to monitor steady breathing but also to detect occasional events, such as apnea.
To achieve this, we first instruct the participant to breathe normally for about one minute.
Then, the participant is asked to hold their breath for about $40$ seconds, and this process is repeated twice.
This approach allows us to capture both normal breathing patterns and apnea events.

\subsubsection{Motion Detection}
\begin{figure}[t]
	\centering
	\subfigure[\label{figLay2} Experimental Layout]{
		\begin{minipage}[t]{0.48\textwidth}	
			\centering
			\includegraphics[width=6.6cm]{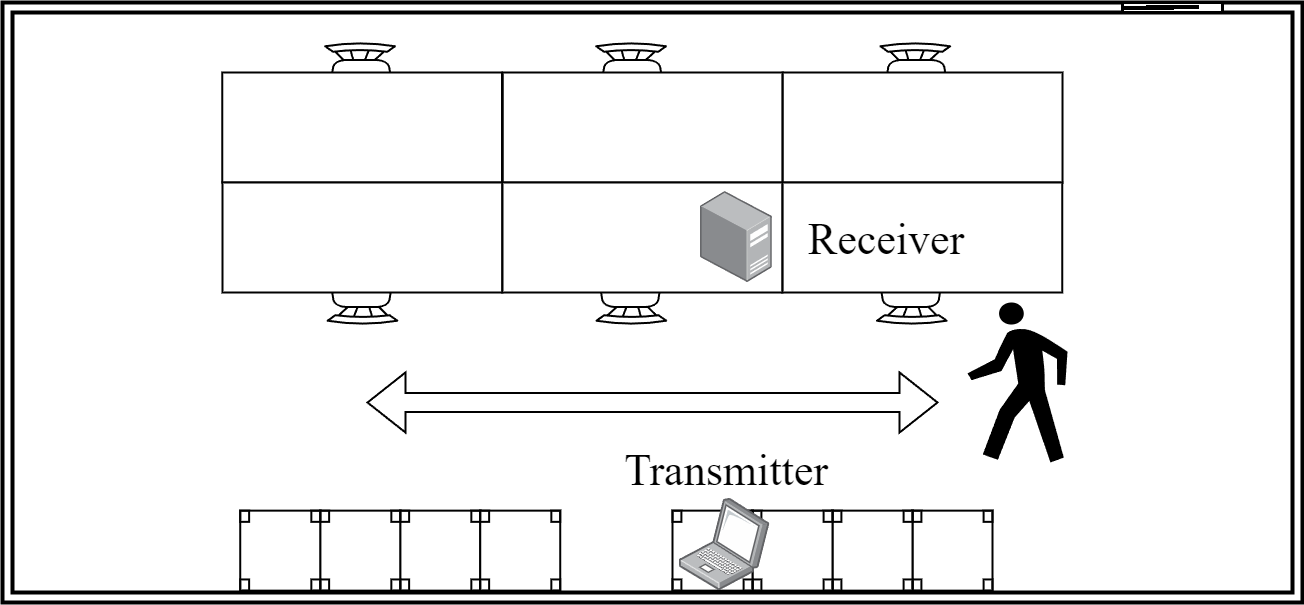}
	\end{minipage}}
	\subfigure[\label{figEnv2} Experimental Environment]{
		\begin{minipage}[t]{0.48\textwidth}
			\centering
			\includegraphics[width=6.6cm]{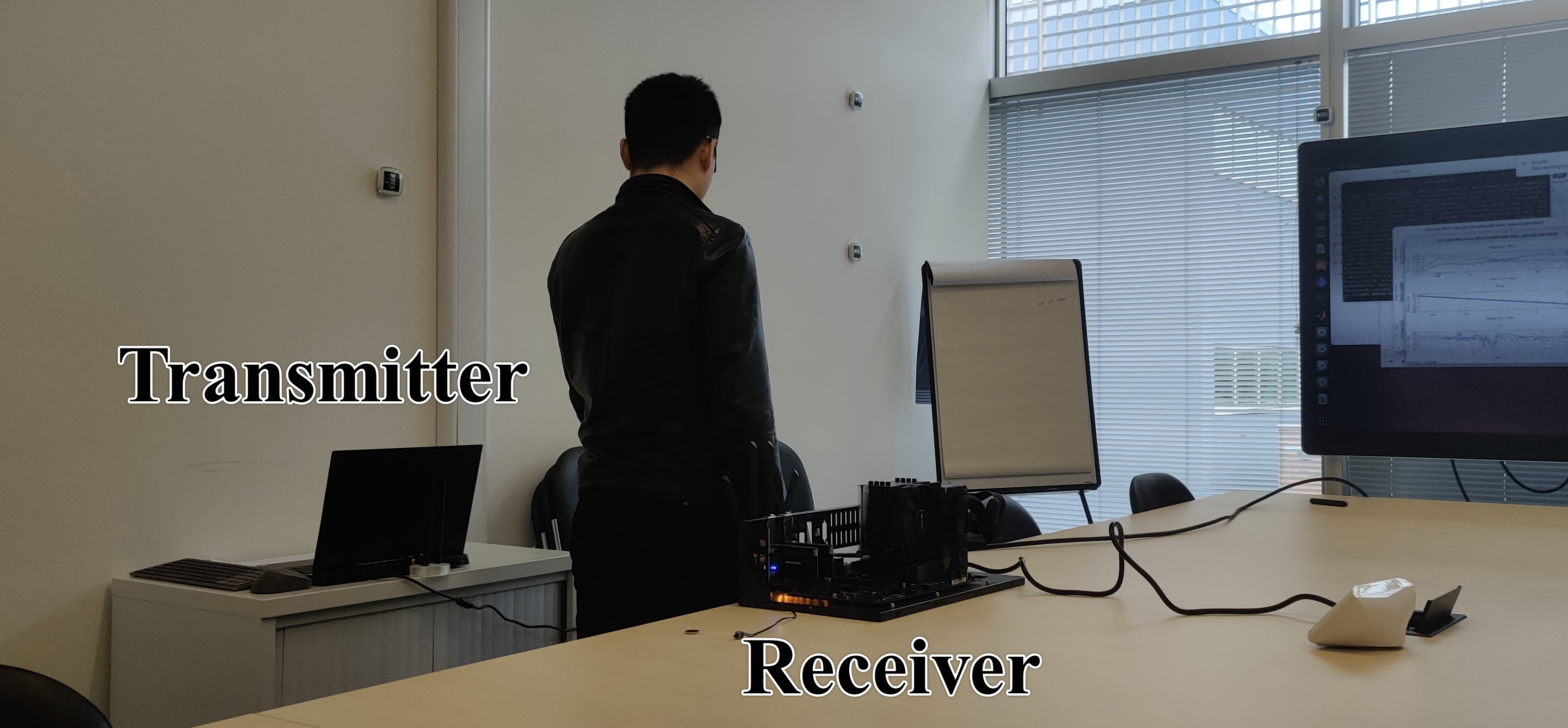}  
	\end{minipage}}
	\caption{\label{figExp2} The experimental layout and environment of motion detection.}
	\vspace{-1em}
\end{figure}

The motion detection experiment aims to capture the movement of a participant within a designated area.
\figref{figLay2} illustrates the experimental layout, which consists of a room equipped with a Wi-Fi \gls{tx} and \gls{rx} placed at opposite ends.
The room is furnished with tables and chairs arranged in a grid-like pattern to simulate a typical indoor environment.
The experimental setup is designed to capture the variations in the wireless signal caused by human movement within the room.
As a person walks between the tables and chairs, their presence and motion alter the signal propagation path between the transmitter and receiver.
By analyzing these signal variations, the system can detect and track human motion in real-time.
\figref{figEnv2} shows the actual experimental environment used in this study.
The experiments were conducted in a conference room or classroom setting, equipped with tables, chairs, and other office furniture.

During the experiments, the participant will walk between the tables and chairs at a normal pace, following a predefined path.
The Wi-Fi signals were continuously monitored and recorded to capture the signal variations caused by the participant's motion.
Overall, this experiment showcases the potential of Wi-Fi sensing for non-invasive, contactless motion detection.
The collected data are then processed through a computationally efficient moving average algorithm, which is described in the next subsection.

\subsection{Moving Average Algorithm}
\begin{figure*}[t]
	\centering
	\subfigure[Respiration Monitoring Using $2.4$ $\mathrm{GHz}$ \gls{wifi} Signals]{\begin{minipage}[t]{0.49\textwidth}	
		\centering
		\includegraphics[width=8.5cm]{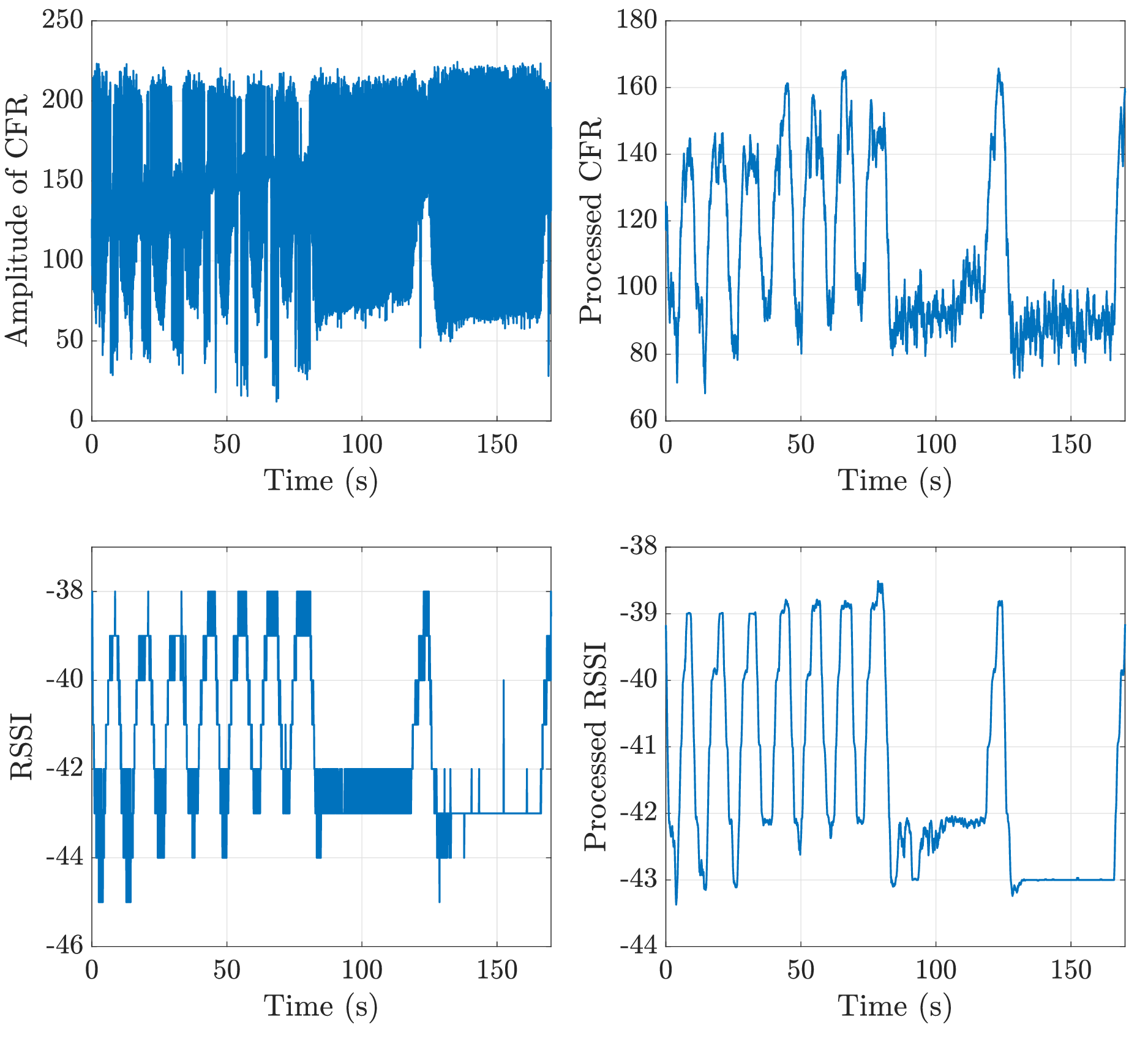}
	\end{minipage}}
	\subfigure[Respiration Monitoring Using $6$ $\mathrm{GHz}$ \gls{wifi} Signals]{\begin{minipage}[t]{0.49\textwidth}
		\centering
		\includegraphics[width=8.5cm]{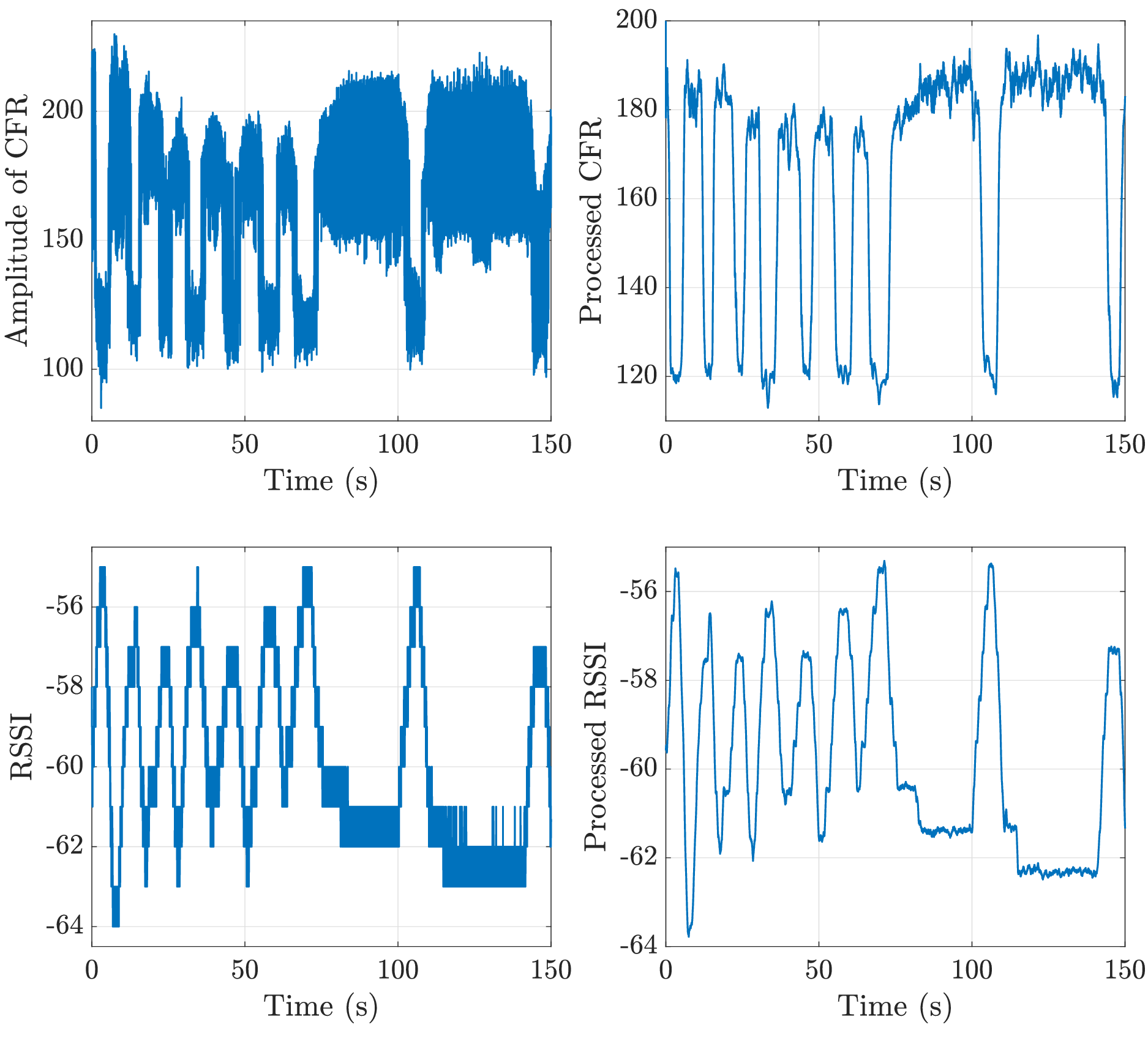}
	\end{minipage}}
	\caption{\label{figEx1} Experimental results of respiration monitoring using Wi-Fi signals in the $2.4$ $\mathrm{GHz}$ and $6$ $\mathrm{GHz}$ frequency bands. The experiments capture both normal breathing patterns and breath-holding periods, demonstrating the effectiveness of using Wi-Fi sensing for non-invasive respiration monitoring. The $6$ $\mathrm{GHz}$ signals provide clearer and more pronounced respiration patterns compared to the $2.4$ $\mathrm{GHz}$ signals.}
	\vspace{-1em}
\end{figure*}
Unlike previous works that convert time-domain signals to frequency-domain signals and apply filtering in the frequency domain, our method is designed to handle non-cyclic human behaviors, particularly in respiration monitoring.
We employ the moving average algorithm, a simple yet effective technique used in signal processing and data analysis, to smooth out short-term fluctuations and highlight longer-term trends or cycles \cite{Hansun2013}.
The moving average algorithm calculates the average of a fixed number of data points in a sliding window that moves through the data series.
As new data points become available, the window slides forward, and the average is recalculated, dropping the oldest data point and including the newest one.
Specifically, let $W$ be the window length, $\gamma_{t}$ be the original data at the time slot $t$, and $\beta_{t}$ be the processed data using the moving average algorithm.
The formula for calculating the moving average is as follows:
\begin{equation} \label{eqn04}
	\beta_{t} = \dfrac{1}{t} \Bigg(\sum_{\tau=1}^{t} \gamma_{\tau}\Bigg), \quad \text{when} \quad t \leq W,
\end{equation}
and
\begin{equation} \label{eqn05}
	\beta_{t} = \dfrac{1}{W} \Bigg(\sum_{\tau=t-W+1}^{t} \gamma_{\tau}\Bigg), \quad \text{when} \quad t > W.
\end{equation}
The moving average algorithm helps to smooth out the data by reducing the impact of short-term fluctuations and highlighting the overall trend.
Note that this is a basic implementation of the moving average algorithm, and there are variations and optimizations that can be applied depending on the specific requirements of the application. 

In our work, we utilize this algorithm to effectively handle non-cyclic human behaviors in respiration monitoring, enabling real-time analysis and detection of respiratory patterns.
By replacing $\gamma_{t}$ with $A_{k,t}$ from \eqref{eqn02} or $I_{t}$ from \eqref{eqn03}, the moving average algorithm can be used to process the \gls{cfr} and \gls{rssi} data, respectively.

\section{Experimental Results} \label{sec04}
This section presents the results of our \gls{wifi} sensing experiments, which consist of two main studies: respiration monitoring and motion detection.
In all the experiments, the window size is set to be $100$.
In the following subsections, we provide a comprehensive demonstration and discussion of each experiment. 

\subsection{Experiment Results of Respiration Monitoring}
\figref{figEx1} shows the results of an experiment using Wi-Fi signals from two different frequency bands for respiratory monitoring.
The figure is divided into two parts: (a) respiration monitoring with $2.4$ $\mathrm{GHz}$ Wi-Fi signals, and (b) respiration monitoring with $6$ $\mathrm{GHz}$ Wi-Fi signals.
In both parts, the top row presents the amplitude of the CFR (left) and the processed CFR using the moving average algorithm (right), while the bottom row depicts the RSSI (left) and the processed RSSI using the moving average algorithm (right).
During the first $40$ seconds, the amplitude of the CFR exhibits periodic fluctuations corresponding to the participant's breathing patterns.
The experiments also capture the $20$-second breath-holding periods, evident from the flat regions in the processed signals.
This demonstrates that \gls{wifi} signals can capture not only cyclically stationary human behavior but also non-stationary behavior.

The moving average algorithm effectively filters out high-frequency noise, resulting in a cleaner and more distinct respiration signal.
Similar to the CFR, the RSSI also captures the periodic variations caused by the participant's respiration, indicating that the easily accessible RSSI data is sufficient for monitoring human respiration behavior.
Comparing the results from the $2.4$ $\mathrm{GHz}$ and $6$ $\mathrm{GHz}$ frequency bands reveals that the $6$ $\mathrm{GHz}$ signals provide clearer and more pronounced respiration patterns.
The higher frequency band offers improved sensitivity to small movements, resulting in more distinct fluctuations in both the CFR and RSSI.
These results demonstrate the effectiveness of noninvasive respiratory monitoring using Wi-Fi sensing, where the moving average algorithm plays a key role in improving signal quality and facilitating accurate respiratory tracking.
The effectiveness of \gls{cfr} and \gls{rssi} analysis of Wi-Fi signals for respiratory monitoring is demonstrated.
In addition to \gls{cfr}, the changes in RSSI are also very clear, providing another method for detecting and monitoring respiratory activity.
Overall, our experimental results successfully demonstrate the feasibility and effectiveness of noninvasive respiratory monitoring using CFR and RSSI analysis of Wi-Fi signals.

\subsection{Experiment Results of Motion Detection}
\begin{figure*}[t]
	\centering
	\subfigure[Motion Detection Using $2.4$ $\mathrm{GHz}$ \gls{wifi} Signals]{\begin{minipage}[t]{0.49\textwidth}
			\centering
			\includegraphics[width=8.5cm]{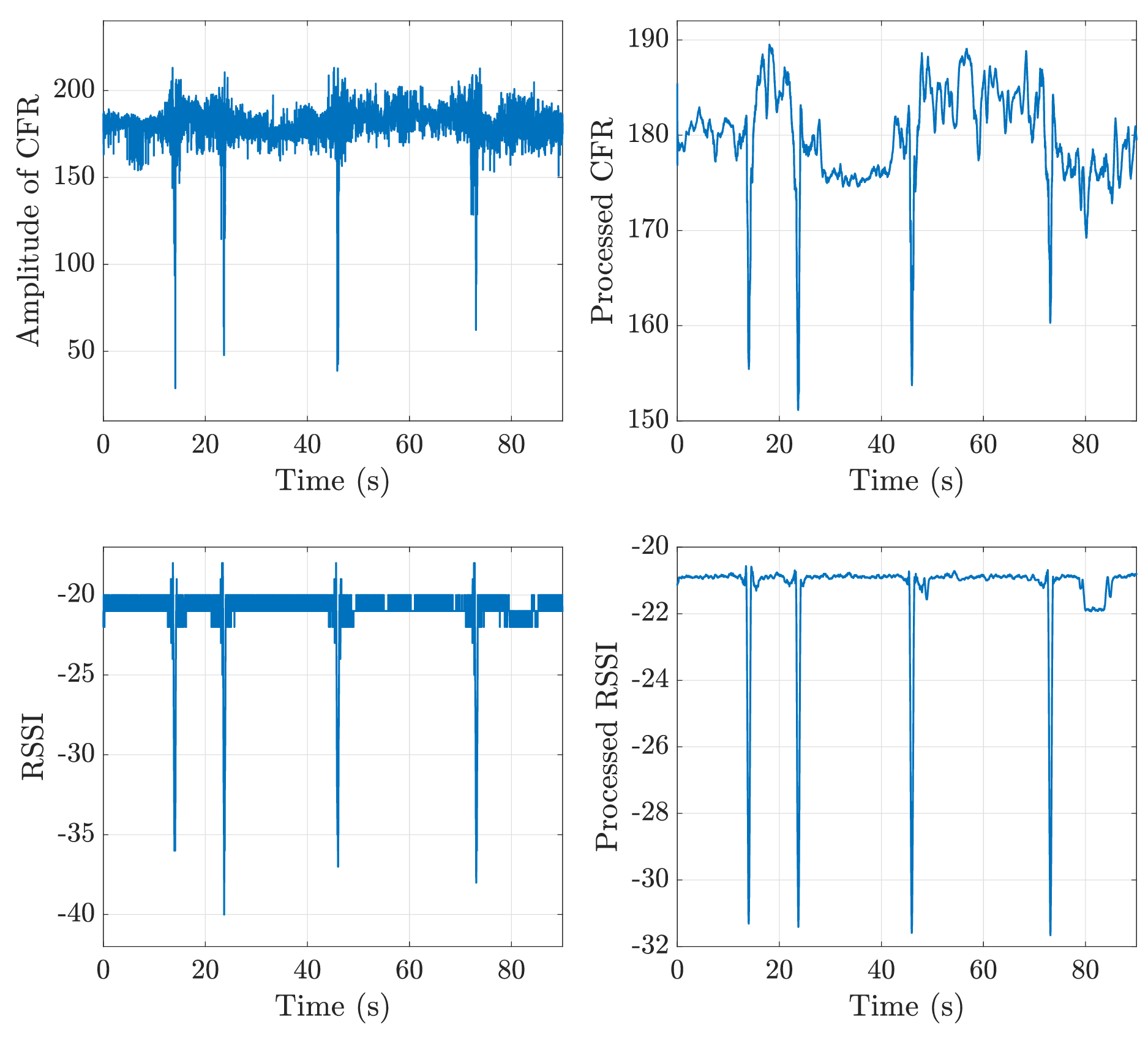}
	\end{minipage}}
	\subfigure[Motion Detection Using $6$ $\mathrm{GHz}$ \gls{wifi} Signals]{\begin{minipage}[t]{0.49\textwidth}
			\centering
			\includegraphics[width=8.5cm]{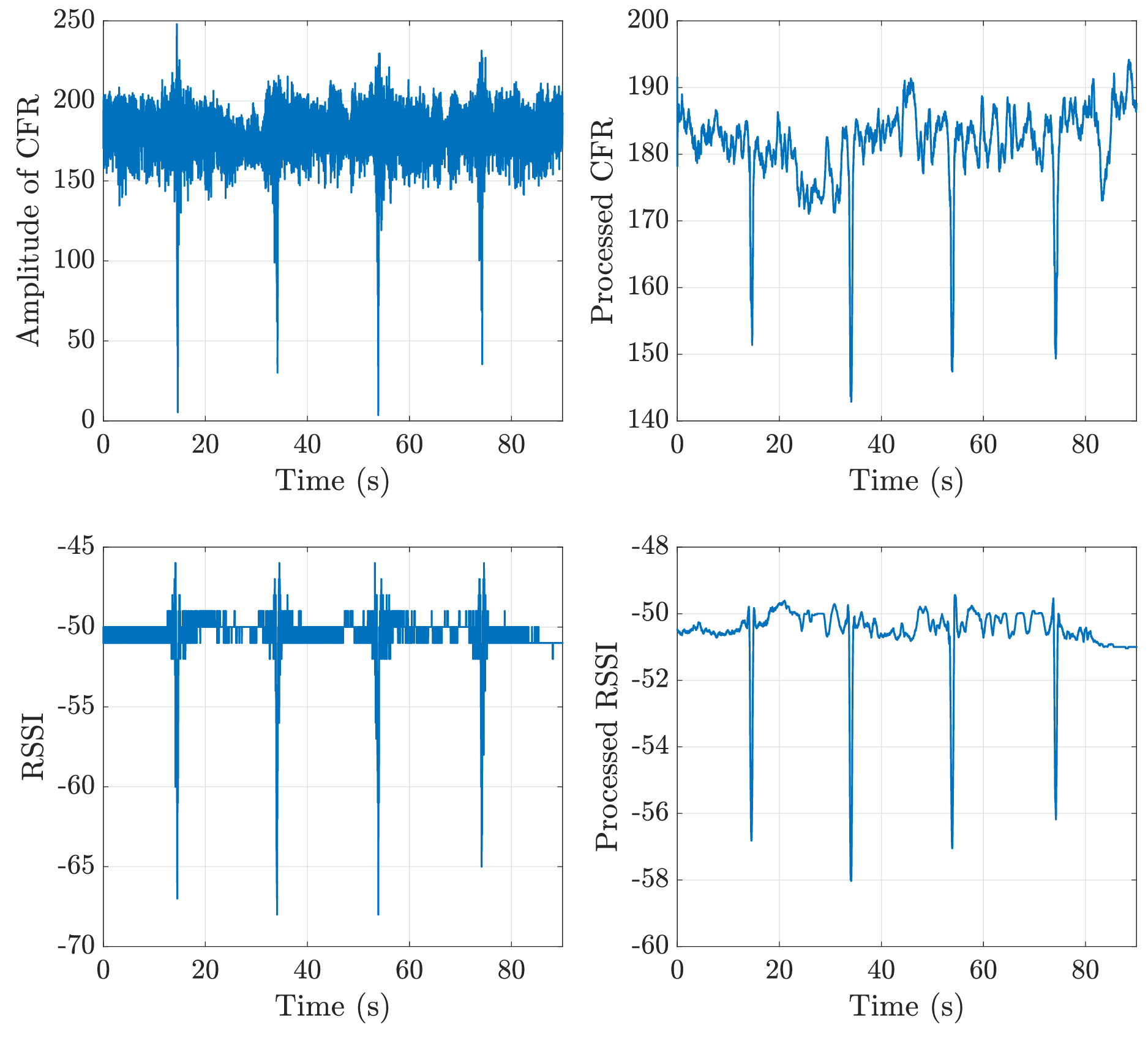}
	\end{minipage}}
	\caption{\label{figEx4} Experimental results of motion detection using Wi-Fi signals in the $2.4$ $\mathrm{GHz}$ and $6$ $\mathrm{GHz}$ bands. The experiments demonstrate the effectiveness of Wi-Fi sensing for non-invasive motion detection, with $6$ $\mathrm{GHz}$ signals providing sharper motion patterns compared to $2.4$ $\mathrm{GHz}$. The processed signals indicate the participant's movement along the predefined path, showcasing the feasibility of Wi-Fi sensing for real-time motion detection.}
	\vspace{-1em}
\end{figure*}

\figref{figEx4} presents the experimental results of motion detection using Wi-Fi signals in two different frequency bands. The figure is divided into two parts: (a) Motion Detection with $2.4$ $\mathrm{GHz}$ Wi-Fi Signals and (b) Motion Detection with $6$ $\mathrm{GHz}$ Wi-Fi Signals.
In both parts, the top row shows the amplitude of the CFR (left) and the processed CFR using the moving average algorithm (right).
As the participant moves within the designated area, the CFR amplitude exhibits significant fluctuations, indicating the presence of motion.
The moving average algorithm effectively smooths out the high-frequency noise, resulting in a clearer representation of the motion patterns.
The bottom row of each part depicts the RSSI (left) and the processed RSSI using the moving average algorithm (right).
Similar to the CFR, the RSSI also captures the variations caused by the participant's movement. 
The moving average algorithm enhances the signal quality, making the motion patterns more discernible.

It is important to note that while both CFR and RSSI can effectively capture motion patterns, RSSI has a significant advantage over CFR in terms of accessibility.
CFR data is only available on specific devices and requires specialized software, limiting its widespread use.
In contrast, RSSI data is readily available on all Wi-Fi devices, making it a more practical and cost-effective option for motion detection applications.
Again, comparing the results from the $2.4$ $\mathrm{GHz}$ and $6$ $\mathrm{GHz}$ frequency bands, we observe that the $6$ $\mathrm{GHz}$ signals provide sharper and more distinct motion patterns.
The higher frequency band offers improved spatial resolution and sensitivity to small movements, resulting in more pronounced fluctuations in both the CFR and RSSI.
Moreover, the processed signals clearly indicate the movement of the participant along the predefined path.
The peaks and valleys in the signals correspond to the participant's proximity to the \gls{tx} and \gls{rx}, respectively. 
This demonstrates that the moving average algorithm plays a vital role in enhancing the signal quality and facilitating real-time motion detection

\section{Conclusion and Outlook}
In this paper, we have investigated the capabilities of Wi-Fi sensing using one of the most advanced \gls{cots} devices, the Intel AX210, for two crucial applications: respiration monitoring and motion detection.
Our experimental results demonstrated the effectiveness of \gls{cfr} and \gls{rssi} in capturing and representing human respiration and motion patterns.
Additionally, we showed that the moving average algorithm enhances signal quality and facilitates accurate tracking of both cyclical and non-cyclical human behaviors.
The findings of this study have significant implications for the development of non-invasive, contactless sensing systems.
Our comparison between $2.4$ $\mathrm{GHz}$ and $6$ $\mathrm{GHz}$ \gls{wifi} signals highlights the benefits of using higher frequency bands.
The use of \gls{cots} devices and the accessibility of \gls{rssi} data make our approach highly practical and cost-effective, paving the way for widespread adoption in various domains, including healthcare, smart homes, security, and the emerging Metaverse applications.

For future work, it is worth noting that higher frequency bands, such as $7$ $\mathrm{GHz}$, $42.5$ $\mathrm{GHz}$, and $71$ $\mathrm{GHz}$, will be utilized in the IEEE 802.11be (Wi-Fi 8) standard \cite{Reshef2022}.
These higher frequencies can further improve the accuracy of \gls{wifi} sensing applications.
Another critical aspect is the integration of Wi-Fi sensing with other sensing modalities, such as vision-based systems or wearable devices, to create multi-modal sensing frameworks.
Furthermore, the privacy and security implications of Wi-Fi sensing need to be carefully considered and addressed.
The development of privacy-preserving algorithms and secure data transmission protocols is essential to ensure the responsible deployment of these technologies in real-world settings.

\section*{Acknowledgement}
This work is supported by the UK Department for Science, Innovation and Technology under the Future Open Networks Research Challenge project TUDOR (Towards Ubiquitous 3D Open Resilient Network). The views expressed are those of the authors and do not necessarily represent the project.

\ifCLASSOPTIONcaptionsoff
\newpage
\fi

\bibliographystyle{IEEEtran}
\bibliography{../IEEEabrv,../thesis_list}
\end{document}